\documentclass[]{llncs}
\usepackage[T1]{fontenc}
\usepackage{graphicx}
\begin{document}
\title{Team LA at SCIDOCA shared task 2025: Citation Discovery via relation-based zero-shot retrieval}
\titlerunning{Team LA at SCIDOCA shared task 2025}
%
\author{Trieu An \inst{1} \and 
Long Nguyen\inst{2} \and
Minh Le Nguyen \inst{1}
}
\authorrunning{An et al.}
%
\institute{Japan Advanced Institute of Science and Technology, Japan \\
\email{\{antrieu,nguyenml\}@jaist.ac.jp}
 \and  Hanoi University of Engineering and Technology, Viet Nam \\
 \email{21025082@uet.vnu.edu.vn}
\\
\email{}}
\maketitle              
\begin{abstract}
The Citation Discovery Shared Task focuses on predicting the correct citation from a given candidate pool for a given paragraph. The main challenges stem from the length of the abstract paragraphs and the high similarity among candidate abstracts, making it difficult to determine the exact paper to cite. To address this, we develop a system that first retrieves the top-k most similar abstracts based on extracted relational features from the given paragraph. From this subset, we leverage a Large Language Model (LLM) to accurately identify the most relevant citation. We evaluate our framework on the training dataset provided by the SCIDOCA 2025 organizers, demonstrating its effectiveness in citation prediction.

\keywords{Citation Discovery  \and Relation extraction \and Text retrival.}
\end{abstract}
\section{Introduction}

Citation prediction is an important task with many practical applications for researchers. When starting to work on a research problem, researchers usually need to survey scientific documents related to that problem. This helps them understand the current state of research, identify limitations in existing work, and develop new ideas to improve those issues. However, with the rapid increase in the number of scientific publications, it is becoming harder to find the exact documents that are truly relevant to their concerns. A research problem can have many aspects, and while there may be multiple approaches to addressing it, not all of them are useful for researchers in their current work \cite{yu2012citation}.

In addition, when writing a scientific document, researchers need to cite the sources that provide them with information, ideas, and viewpoints. This process requires precision to ensure that readers can trace and verify the information mentioned in the document. A lack of accurate citation can lead to confusion or misinterpretation, which affects the credibility of the research \cite{wang2019attention}. Therefore, a system that helps researchers search for a set of documents closely related to the paragraph they are working on would be very useful.

In this report, we present our system for citation discovery in SCIDOCA 2025 Shared Task 1. Our system leverages a Large Language Model (LLM) to extract relation triples at the document level from a given paragraph and retrieve the top-k most relevant documents. From these retrieved documents, we designed a set of prompts to identify the correct citation for the given paragraph.

To evaluate the effectiveness of our system, we compared it with baseline methods such as TF-IDF and Dense Vector Retrieval (DVR) on the dataset provided by SCIDOCA 2025 Shared Task 1. Our results show positive improvements in F1-score, demonstrating that integrating relation extraction with LLM-based citation retrieval is an effective approach for this task.
\section{Related works}
\subsection{Text Retrieval}

Text retrieval has been a fundamental problem in information retrieval (IR), evolving from traditional keyword-based approaches to complicated deep learning methods. Over the years, researchers have explored various techniques, including term-based retrieval, dense vector representations, and neural-based retrieval systems.\\

\textbf{Traditional Term-Based Retrieval}\\

One of the earliest and most widely used approaches to text retrieval is Term Frequency-Inverse Document Frequency (TF-IDF), which ranks documents based on keyword importance \cite{salton1988term}. The Vector Space Model (VSM) \cite{salton1975vector} represents documents as high-dimensional vectors, allowing for similarity-based retrieval using cosine similarity. Additionally, BM25 \cite{robertson1994some} introduced probabilistic ranking by considering term frequency, document length, and inverse document frequency, significantly improving retrieval performance. These methods remain widely used due to their efficiency and interpretability.\\

\textbf{Dense Vector-Based Retrieval}\\

As text retrieval tasks became more complex, researchers began exploring dense vector representations. Word embeddings, such as Word2Vec \cite{mikolov2013distributed}, GloVe \cite{pennington2014glove}, and FastText \cite{bojanowski2017enriching}, improved text retrieval by capturing semantic relationships between words. However, these static embeddings struggled with contextual variations. The introduction of transformer-based embeddings, such as BERT \cite{devlin2019bert} and RoBERTa \cite{liu2019roberta}, enabled Dense Passage Retrieval (DPR) \cite{karpukhin2020dense}, which uses bi-encoder models to encode queries and documents into dense vectors for similarity retrieval. These methods significantly improved retrieval accuracy compared to traditional term-based approaches.\\

\textbf{Neural Retrieval and Re-Ranking}\\

End-to-end neural retrieval models, such as T5-based retrieval models \cite{ni2021sentence} and Gated Transformer Retrieval (GTR) \cite{ni2021gtr}, fine-tune transformer models for document retrieval tasks. Additionally, retrieval models benefit from re-ranking mechanisms, where a first-pass retrieval model generates candidate documents, followed by a cross-encoder that reranks documents based on deeper contextual understanding \cite{nogueira2019passage}. Re-ranking techniques significantly enhance retrieval precision, particularly in open-domain question answering and legal text retrieval.\\

\textbf{Retrieval-Augmented Generation (RAG)}\\

Integrating Large Language Models (LLMs) into text retrieval systems has led to significant advancements in how information is accessed and processed. A notable development in this area is Retrieval-Augmented Generation (RAG), a technique that combines the strengths of retrieval systems with the generative capabilities of LLMs.\cite{gao2023retrieval} This approach enhances the ability of LLMs to access and incorporate external information, thereby improving their performance on knowledge-intensive tasks such as Text Retrieval.

\subsection{Document-Level Relation Extraction}
Traditional relation extraction models focus on sentence-level relations, but many real-world applications require document-level relation extraction (DocRE). In this setting, relations are extracted across multiple sentences, making the task significantly more challenging due to coreference resolution and entity linking \cite{yao2019docred}. Graph-based models and multi-hop reasoning have been proposed to address these challenges by incorporating global document context \cite{christopoulou2019connecting}.

\subsection{Task 1 SCIDOCA 2025 shared task}
The objective of Task 1 of the SCIDOCA shared task is to predict relevant citations for a given paragraph without specifying the exact sentence where the citation belongs. The input consists of a paragraph and a set of candidate abstracts, and the system must identify the ID of the correct abstract from the candidate set. This task presents several challenges. First, both the query paragraph and the candidate abstracts are long, introducing significant noise in the retrieval process. Additionally, the candidate set is pre-selected by the organizers to include distracting abstracts, making it more difficult to identify the correct citation. As a result, effective retrieval strategies must handle long-form text while distinguishing subtle differences between highly similar abstracts. The performance of each competing system is evaluated using the F1 score, ensuring that models effectively balance precision and recall in citation prediction.
\section{Methods}
\begin{figure}[h]
    \centering
    \includegraphics[width=0.8\textwidth]{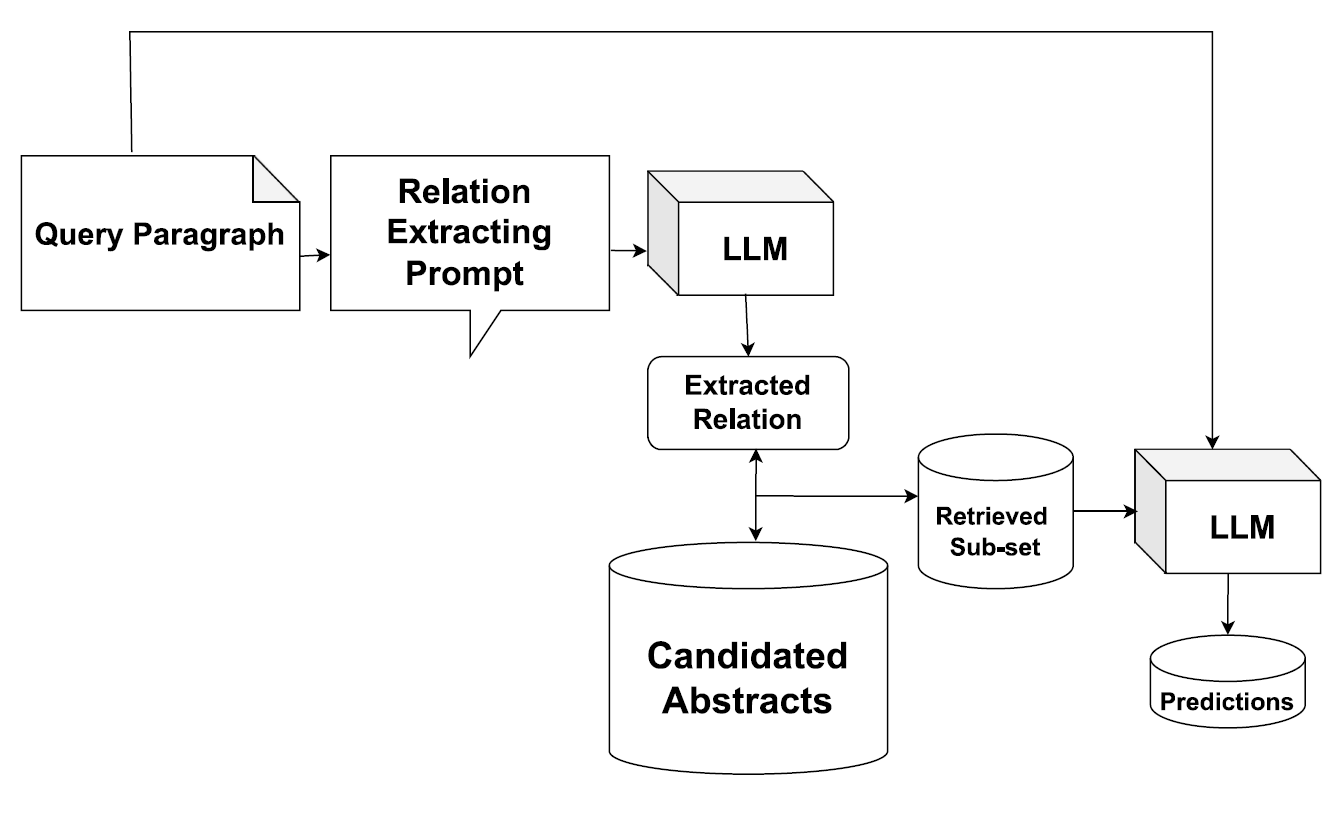}
    \caption{Overview of our system.}
    \label{fig:method}
\end{figure}

In this section, we present our relation-enhanced retrieval approach for citation prediction in document-level relation extraction. The method is designed to improve citation accuracy by leveraging relational information extracted from the query paragraph. The overall process is illustrated in Figure~\ref{fig:method}.

The process begins with a query paragraph, which serves as the context for citation retrieval. Since the paragraph may contain multiple concepts and research directions, it is first processed using a relation-extracting prompt. This prompt instructs a Large Language Model (LLM) to extract structured relational triples or key concepts that capture the essential connections between entities present in the text. By structuring the input in this way, the system ensures that the retrieval process is guided by meaningful semantic relationships rather than simple keyword matching.

Once the relational information is extracted, it is used to retrieve a relevant subset of abstracts from a predefined candidate pool. This retrieval step is critical because the candidate set is designed to include distracting abstracts that may contain overlapping content but are not the correct citation. By filtering the pool based on the extracted relations, the system significantly reduces noise and improves the likelihood of retrieving the correct reference.

After retrieving the most relevant subset of abstracts, a second LLM is employed to make the final citation prediction. This LLM compares the retrieved abstracts with the query paragraph, considering both textual similarity and relational coherence. The model then outputs the ID of the most relevant abstract, which serves as the final predicted citation.

This two-step approach, combining relation extraction with a retrieval and ranking mechanism, improves citation prediction by ensuring that the system focuses on meaningful contextual relations. By reducing reliance on surface-level text similarity, it enhances the ability to distinguish between highly similar abstracts. The system's performance is evaluated using the F1 score, which balances precision and recall to ensure robust citation predictions.

\section{Experiments}
To evaluate the performance of our system, we conduct experiments on 1,000 randomly selected queries from the dataset. The F1 score is used as the evaluation metric, and we compare our method against baseline approaches utilizing TF-IDF and Dense Vector Retrieval. During experiments, we chose the model "mistralai/Mistral-7B-Instruct-v0.3" for relation extracting and for retrieving the correct citation as well. For the first retrieving step we collect the top 20 most related documents from the candidate documents pool. 

The experimental results presented in Tables~\ref{tab:retrieval_comparison} and~\ref{tab:llm_comparison} provide insights into the effectiveness of different retrieval methods and their impact on citation prediction performance.

From Table~\ref{tab:retrieval_comparison}, we observe that traditional retrieval methods such as TF-IDF and Dense Retrieval achieve high recall scores, but their precision is relatively low. This indicates that while these methods successfully retrieve the correct document in many cases, they also return a large number of irrelevant documents, thereby reducing precision. As expected, increasing the number of retrieved documents (top-k) leads to an increase in recall but at the cost of further declining precision. The best performing traditional retrieval method, dense retrieval-10, achieves an F1 score of 0.3354, outperforming TF-IDF-based methods.

Our relation-based retrieval approach, on the other hand, demonstrates a substantial improvement in precision (e.g., 0.8506 for top-10). However, this comes at the cost of reduced recall, suggesting that relation extraction filters out many irrelevant documents but may also discard some relevant ones. This trade-off is evident when comparing our method to dense retrieval: while relation-based retrieval provides a more precise set of candidate documents, it does not necessarily maximize recall.

Moving to Table~\ref{tab:llm_comparison}, where we analyze LLM-based inference, we note that integrating LLMs with retrieval significantly enhances recall compared to raw retrieval scores. The LLM inference with relation-based retrieval achieves an F1 score of 0.2912, showing a balanced performance similar to dense retrieval but with a different recall-precision trade-off. Notably, the combination of LLM with TF-IDF leads to slightly better precision than LLM with dense retrieval, but its recall remains lower.

These findings highlight the importance of relation-based retrieval in refining candidate documents before passing them to the LLM, ensuring that the final prediction is not only relevant but also accurate. The trade-off between precision and recall is a key consideration in designing an effective citation discovery system, and our results indicate that an optimal combination of dense retrieval for recall and relation-based filtering for precision may yield the best performance.

\begin{table}[h]
    \centering
    \renewcommand{\arraystretch}{1.2} 
    \begin{tabular}{|c|c|c|c|}
        \hline
        \textbf{} & \textbf{Recall} & \textbf{Precision} & \textbf{F1 Score} \\
        \hline
        TF-IDF-10 & 0.8259 & 0.2060 & 0.3297 \\
        \hline
        TF-IDF-15 & 0.9115 & 0.1715 & 0.2886 \\
        \hline
        TF-IDF-20 & 0.9533 & 0.1534 & 0.2643 \\
        \hline
        dense retrieval-10 & 0.8401 & 0.2095 & 0.3354 \\
        \hline
        dense retrieval-15 & 0.9215 & 0.1733 & 0.2918 \\
        \hline
        dense retrieval-20 & 0.9615 & 0.1547 & 0.2665 \\
        \hline
        relation-based-10 & 0.8506 & 0.2156 &  0.344 \\
        \hline
        relation-based-15 & 0.93 & 0.1772 & 0.2976 \\
        \hline
        relation-based-20 & 0.967 & 0.1576 & 0.2711 \\
        \hline
    \end{tabular}
    \caption{Performance comparison of different retrieval methods. Each method is implemented with the most relevant documents from the different top-k.}
    \label{tab:retrieval_comparison}
\end{table}

\begin{table}[h]
    \centering
    \renewcommand{\arraystretch}{1.2} 
    \begin{tabular}{|l|c|c|c|}
        \hline
        & \textbf{Recall} & \textbf{Precision} & \textbf{F1 Score} \\
        \hline
        LLM inference with TF-IDF & 0.4079 & 0.2347 & 0.2980 \\
        \hline
        LLM inference with Dense retrieval & 0.3253 & 0.2590 & 0.2884 \\
        \hline
        LLM inference with relation-base & 0.4626 & 0.2125 & 0.2912 \\
        \hline
    \end{tabular}
    \caption{Performance comparison of LLM inference methods.}
    \label{tab:llm_comparison}
\end{table}

\section{Conclusion}

In this paper, we presented a relation-based retrieval approach for citation discovery, leveraging document-level relation extraction and large language models (LLMs) to improve citation prediction accuracy. Our system extracts key relational features from the query paragraph to retrieve a more semantically meaningful subset of candidate abstracts, reducing the noise introduced by traditional text-based retrieval techniques.

Experimental results demonstrate that while dense vector retrieval achieves high recall, its precision remains relatively low. In contrast, our relation-based retrieval significantly improves precision, ensuring that only the most relevant abstracts are considered. Integrating LLM inference further enhances recall, leading to a more balanced retrieval and ranking strategy. The combination of relation-based retrieval with LLM inference proves to be an effective approach for citation prediction tasks.

For future work, we aim to explore hybrid retrieval models that dynamically adjust the recall-precision trade-off based on paragraph characteristics. Additionally, incorporating graph-based relational reasoning could further improve retrieval effectiveness by capturing deeper contextual dependencies across documents.

%
%
%
\bibliographystyle{splncs04}
\bibliography{ref}
%

\end{document}